\documentclass[a4paper,11pt]{article}
\usepackage{pos}
\usepackage{orcidlink}
\usepackage{subfig}
\usepackage{diagbox}
\usepackage{makecell, multirow}
\usepackage{pgf}
\usepackage{soul}
\usepackage{graphicx, color}
\usepackage{rotating}
\allowdisplaybreaks

\definecolor{link_blue}{RGB}{51,102,204}
\hypersetup{%
pdftitle = {title},
pdfsubject = {},
pdfkeywords = {},
colorlinks = {true},
filecolor = {black},
linkcolor = {link_blue},
menucolor = {black},
citecolor = {link_blue},
urlcolor = {link_blue},
}{}

\title{Precisely determining the ground state mass of Spin-3/2 $\Omega_{ccc}$ baryon from Lattice QCD}
\ShortTitle{Precisely determining the ground state mass of Spin-3/2 $\Omega_{ccc}$}

\author*[a]{Navdeep Singh Dhindsa~\orcidlink{0000-0002-3133-6979}}
\author[a]{Debsubhra Chakraborty~\orcidlink{0000-0001-5815-4182}}
\author[a]{Archana Radhakrishnan~\orcidlink{0000-0001-9357-1360}}
\author[a]{Nilmani Mathur~\orcidlink{0000-0003-2422-7317}}
\author[b,c]{M.~Padmanath~\orcidlink{0000-0001-6877-7578}}

\affiliation[a]{Department of Theoretical Physics, Tata Institute of Fundamental Research,\\Homi Bhabha Road, Mumbai 400005, India}
\affiliation[b]{The Institute of Mathematical Sciences,\\CIT Campus, Chennai, 600113, India}
\affiliation[c]{Homi Bhabha National Institute,\\Training School Complex, Anushaktinagar, Mumbai 400094, India}

\emailAdd{navdeep@theory.tifr.res.in}
\abstract{We present the most precise determination to date of the ground-state masses of the triply charmed baryons with both parities, obtained by continuum extrapolation and fully addressing the systematic uncertainties. The calculations are performed on six $N_f=2+1+1$ HISQ ensembles, generated by the MILC collaboration, with two complementary setups for the valence charm action, one using the HISQ action and the other using the overlap fermion action. Our prediction for the mass of the lowest two triply charmed spin-3/2 baryons are: $M_{\Omega_{ccc}} (3/2^{+}) =  4793 (5) \left(^{+11}_{-8}\right)$ MeV, and $M_{\Omega_{ccc}} (3/2^{-}) = 5094 (12) \left(^{+19}_{-17}\right)$ MeV.}

\FullConference{The 42nd International Symposium on Lattice Field Theory (LATTICE2025)\\
2-8 November 2025\\
Tata Institute of Fundamental Research, Mumbai, India\\}

\begin{document}
\maketitle
\section{\label{sec:precise}Precise $\Omega_{ccc}(3/2^{\pm})$ Masses}
\begin{figure}[h!]
    \centering
    \begin{minipage}[b]{0.95\textwidth} 
        \centering
        \resizebox{\linewidth}{!}{\input{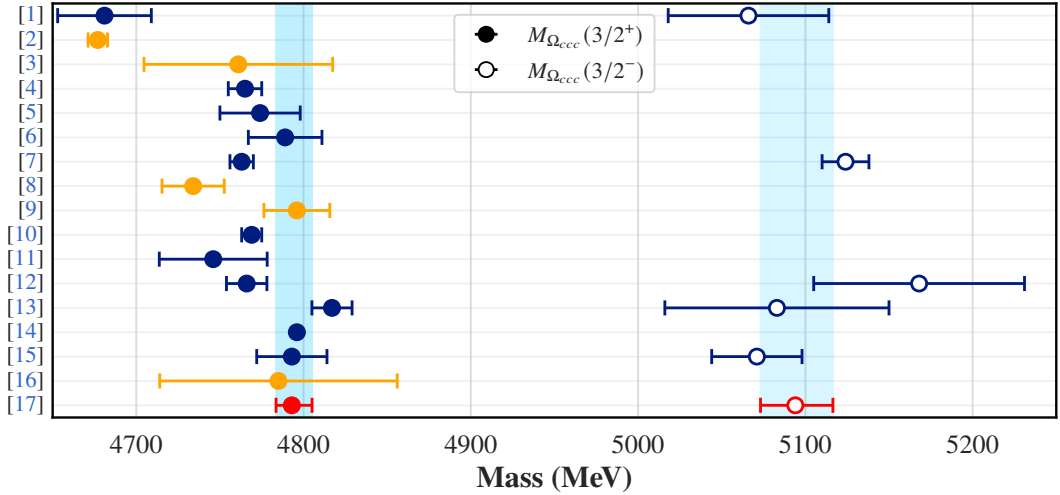}} 
    \end{minipage}
    \caption{Summary of results for the ground-state masses of the $\Omega_{ccc}(3/2^{+})$ and $\Omega_{ccc}(3/2^{-})$ baryons from various lattice calculations. Results that include a continuum extrapolation are shown as orange circles, while those obtained at a single lattice spacing are indicated by blue circles. Filled circles correspond to the $3/2^{+}$ state, whereas open circles denote the $3/2^{-}$ state. Our continuum extrapolated values are highlighted in red, and the accompanying blue band is drawn through the red point to aid visual comparison with the other determinations.}
    \label{fig:main}
\end{figure}

In this talk, we present our recent lattice QCD calculation of the triply charm baryon $\Omega_{ccc}(3/2)$ ground state masses, which is the most precise determination to date. Heavy-flavor baryons, in particular the $\Omega_{ccc}(3/2)$-baryon, provide promising handles to the physics of confinement that are obscure in the light hadrons and in heavy meson systems. Phenomenological studies and lattice investigations offer valuable insight towards theoretical understanding of these systems and serve as a guide for the ongoing and future experimental searches. However, on the phenomenological front, predictions for $\Omega_{ccc}$-baryon mass span a wide range $400\ \text{MeV}$, whereas the existing lattice QCD results are spread over an energy interval of $100\ \text{MeV}$ for the $\Omega_{ccc}(3/2^{+})$ state \cite{Chiu:2005zc, Alexandrou:2012xk, Briceno:2012wt, Basak:2012py, Durr:2012dw, PACS-CS:2013vie, Padmanath:2013zfa, Alexandrou:2014sha, Brown:2014ena, Can:2015exa, Alexandrou:2017xwd, Chen:2017kxr, Bahtiyar:2020uuj, Lyu:2021qsh, Li:2022vbc, Alexandrou:2023dlu}, with an even larger variation for the negative-parity channel. In this calculation, we perform a precise and systematically controlled lattice QCD determination of the $\Omega_{ccc}$ baryon masses with both parities \cite{Dhindsa:2024erk} that can provide a controlled benchmark in future experimental searches. Our final results for the $\Omega_{ccc}(3/2)$-baryon ground state masses with either parity are:
\begin{eqnarray}
M_{\Omega_{ccc}} (3/2^{+}) &=&  4793 (5) \left(^{+11}_{-8}\right)\mbox{ MeV  and}\nonumber \\
M_{\Omega_{ccc}} (3/2^{-}) &=& 5094 (12) \left(^{+19}_{-17}\right)\mbox{ MeV.}
\end{eqnarray}
Following a brief survey on the single flavored baryons and the $\Omega_{ccc}$-baryons in Sec. \ref{sec:intro}, we present our research methodology in Sec.~\ref{sec:setup}. In Fig.~\ref{fig:main}, we present a comparison of our results with previous lattice determinations. Additionally, we provide a comparative summary on various technical details involved in different lattice determinations in Table~\ref{tab:lat_reslt_sum}. 

\begin{sidewaystable*}[htp]
\setlength{\tabcolsep}{2.1mm}{
\centering
\renewcommand\arraystretch{1.2} 
\begin{tabular}{|c|c|c|c|c|c|c|c|c|}
\hline 
\hline 
Ref. (Year)      &$N_{f}$           &$a(fm)$         &$m_{\pi}$ (MeV)    &$S_{q}^{sea}$       &$S_{c}^{val}$     & Continuum &$M_{\Omega_{ccc}}(\frac{3}{2}^{+})$  &$M_{\Omega_{ccc}}(\frac{3}{2}^{-})$ 
\\
 &          &        &    & && Extrapolation & (MeV) &(MeV)
\\

\hline
\cite{Chiu:2005zc} (2005)     &quench   &0.0882      &-    & Wilson    &DW    & $\textcolor{blue!50!black!100}{\textbf{No}}$&4681(28)      &5066(48)   \\
\cite{Alexandrou:2012xk} (2012)     & 2,2+1 &   0.056-0.089   &  260-470  &  TM  & OS  & $\textcolor{orange!80!yellow!100}{\textbf{Yes}}$ (3) &4677(5)(3)    & -  \\
\cite{Briceno:2012wt} (2012)    &2+1+1       &0.06-0.12     &220-310    &HISQ           &RHQA &  $\textcolor{orange!80!yellow!100}{\textbf{Yes}}$ (3)   &4761(52)(21)(6)    &-   \\
\cite{Basak:2012py} (2012) &  2+1+1     &  0.06-0.09   &  316-329  &     HISQ      &  Overlap   &  $\textcolor{blue!50!black!100}{\textbf{No}}$ &4765(10)  &-   \\
\cite{Durr:2012dw} (2012) & 2     &  0.0728 & 280 & Clover  &   Brillouin        & $\textcolor{blue!50!black!100}{\textbf{No}}$  &4774(24)    & -\\
\cite{PACS-CS:2013vie} (2013) &  2+1    & 0.0899  & 135  & Clover &    RHQA       & $\textcolor{blue!50!black!100}{\textbf{No}}$ & 4789(22)    & - \\
\cite{Padmanath:2013zfa} (2013) &  2+1    & 0.0351  & 390 & Clover  &    Clover       & $\textcolor{blue!50!black!100}{\textbf{No}}$ & 4763(7)     & 5124(14)\\
\cite{Alexandrou:2014sha} (2014) &  2+1+1    &  0.065-0.094 & 210-430 & TM &  OS         &  $\textcolor{orange!80!yellow!100}{\textbf{Yes}}$ (3)&4734(12)(11)(9)     & -\\
\cite{Brown:2014ena} (2014) &  2+1    &  0.085-0.11  & 227-419 & DW &   RHQA        & $\textcolor{orange!80!yellow!100}{\textbf{Yes}}$ (2)  & 4796(8)(18)    & -\\
\cite{Can:2015exa} (2015) &  2+1    &  0.0907  & 156 & Wilson &  Clover         &  $\textcolor{blue!50!black!100}{\textbf{No}}$  & 4769(6)   & - \\
\cite{Alexandrou:2017xwd} (2017) & 2     & 0.0938  & 130 & TM Clover &  OS         & $\textcolor{blue!50!black!100}{\textbf{No}}$ &4746(4)(32)     & -\\
\cite{Chen:2017kxr} (2017) &  2+1+1    &  0.063 &  280 & DW &   DW    &  $\textcolor{blue!50!black!100}{\textbf{No}}$  & 4766(5)(11)      & 5168(37)(51)\\ 
\cite{Bahtiyar:2020uuj} (2020) &  2+1    &  0.0907 & 156 & Clover & Clover      &  $\textcolor{blue!50!black!100}{\textbf{No}}$  &  4817(12)     & 5083(67)\\
\cite{Lyu:2021qsh} (2021) &   2+1   & 0.0846  & 146 & Clover &  RHQA     &  $\textcolor{blue!50!black!100}{\textbf{No}}$  &   4796(1)    & -\\
\cite{Li:2022vbc} (2022) &  2+1     & 0.0711-0.0828  & 278-300 & DW &  Overlap     &  $\textcolor{blue!50!black!100}{\textbf{No}}$  &  4793(21)     & 5071(27)\\
\cite{Alexandrou:2023dlu} (2023) & 2+1+1     & 0.057-0.080  & 137-141 & TM Clover &  OS     &   $\textcolor{orange!80!yellow!100}{\textbf{Yes}}$ (3) &    4785(71)   & -\\
\cite{Dhindsa:2024erk} This work &  2+1+1    & 0.0327-0.1207  & 216-329 \cite{PhysRevD.98.074512} & HISQ & HISQ, Overlap      &  $\textcolor{red}{\textbf{Yes}}$ (5)  &   4793(5)($^{+11}_{-8}$)    & 5094(12)($^{+19}_{-17}$)\\
\hline \hline
\end{tabular}}
\caption{A summary of existing lattice QCD determinations of the $\Omega_{ccc}(3/2^{+})$ and $\Omega_{ccc}(3/2^{-})$ masses is presented. The entry from the present study appears in the final row. For each calculation, we list key lattice parameters, including the number of dynamical flavors ($N_{f}$), lattice spacing ($a$), pion mass ($m_{\pi}$), and the actions employed for the sea ($S^{\text{sea}}_{q}$) and valence charm quarks ($S^{\text{val}}_{c}$). We also indicate whether a continuum extrapolation was performed; when applicable, the number of lattice spacings used in the extrapolation is shown in parentheses. Abbreviations used are: HISQ (highly improved staggered quark), DW (domain-wall), TW (twisted mass), OS (Osterwalder–Seiler), and RHQA (relativistic heavy-quark action). Reported masses are rounded to the nearest integer. For calculations performed at multiple lattice spacings without a continuum limit, the values from the finest lattice spacing are quoted.}\label{tab:lat_reslt_sum}
\end{sidewaystable*}

\section{\label{sec:intro}Motivation}

Assuming only a single quark flavor, the strongly interacting world would be a simplified one with no light nucleon, and the lightest baryon being the maximally symmetric spin–$3/2$ state. The real QCD contains two light flavors, but the underlying expectation of a maximally symmetric ground state survives in the strange sector through the well-established $\Omega(sss)$ baryon with $J^{P}=3/2^{+}$ \cite{Barnes:1964pd,BaBar:2006omx}. The discovery of the $\Omega(sss)$ baryon with $J^{P}=3/2^{+}$ played a central role in establishing the quark model and in revealing the color degree of freedom. Naturally, we expect that the charm sector should host its own fully symmetric triply charmed baryon, the $\Omega_{ccc}$, whose existence is a natural consequence of QCD even though it has not yet been observed experimentally. The $\Omega_{QQQ}$ ($Q=c,b$) baryons, analogous to heavy quarkonia, offer a simplistic system for studying quark-quark interactions and quark confinement, without the complexities from valence light quark dynamics \cite{Meinel:2012qz}. As highlighted by Bjorken, the structural simplicity of triply heavy baryons makes them especially suited for investigating parton-level dynamics within baryons \cite{Bjorken:1985ei}.

Recent developments at experimental facilities further strengthen the motivation for a refined theoretical investigation of this baryon. The discovery of the doubly charmed $\Xi_{cc}^+$ baryon \cite{LHCb:2017iph} and the observation of several excited $\Omega_{c}^{0}$ resonances at experimental facilities \cite{LHCb:2017uwr} provide strong motivation for further exploration of multi-charm systems. As the experimental facilities move toward higher luminosities, many additional heavy hadrons are anticipated to be discovered, with the $\Omega_{ccc}$ expected to come within experimental reach. Its strong decay is forbidden, making it potentially long-lived and therefore challenging to detect directly unless the lighter hadrons to which they decay are well understood. But recent observations of rare multi-charm signatures, including tri-$J/\psi$ events \cite{CMS:2021qsn}, indicate that current experiments are entering the regime where triply charmed baryons could be observed in the near future. A controlled, continuum-extrapolated first principles determination of the $\Omega_{ccc}$-baryon mass is therefore essential to provide a robust benchmark for forthcoming experimental searches. In the work presented here \cite{Dhindsa:2024erk}, we perform such a determination using lattice QCD simulations for both the positive- and negative-parity $\Omega_{ccc}$ states by utilizing six ensembles, one of which features the finest lattice spacing employed to date for this system, together with multiple valence actions. The technical setup and results are discussed in the next section. 

\section{\label{sec:setup}Numerical Setup and Results}
We use six $N_f = 2+1+1$ lattice QCD ensembles with Highly Improved Staggered Quark (HISQ) sea-quark dynamics generated by the MILC Collaboration \cite{Follana:2006rc, MILC:2012znn, PhysRevD.98.074512}, spanning two spatial volumes and five lattice spacings, which together provide the basis for our continuum analyses\footnote{Prior to this, only five lattice studies have performed continuum extrapolations for the $3/2^+$ state, yet the extracted masses vary substantially due to the differing approaches employed (see Table~\ref{tab:lat_reslt_sum} for details), and no continuum-extrapolated result exists for the $3/2^-$ channel.}. On the valence side, we employ two actions, Overlap and HISQ, to ensure robustness against action-dependent systematics. A detailed summary of all ensembles and simulation parameters is provided in Table~\ref{tab:Lattice_details_ov_hisq}. A comprehensive description of the bare-quark mass tuning procedure and the lattice-spacing determinations used for these ensembles is available in Ref.~\cite{Dhindsa:2024erk}.

\begin{table}[h!]
    \centering
    \renewcommand\arraystretch{1.2}  
    \addtolength{\tabcolsep}{1pt}    
    \begin{tabular}{|c|c|c|c|c|c|}
     \hline\hline 
      Lattice Size & \multicolumn{2}{c|}{Lattice Spacing ($a$ fm)} & \multirow{2}{*}{$M_{\pi}^{sea}$ (MeV)\cite{PhysRevD.98.074512}} & \multicolumn{2}{c|}{\(n_{\text{meas}}\)} \\ \cline{2-3} \cline{5-6}
      $N_s^3\times N_t$ & Overlap \cite{MILC:2012znn,MILC:2015tqx,PhysRevD.98.074512} & HISQ \cite{HPQCD_2020} & & Overlap & HISQ   \\ 
     \hline    
      $24^3 \times 64$ & $0.1207 (11)$ & $0.12404(67)$ & $305$ & $294$ & $1400$ \\
      $32^3 \times 96$ & $0.0888 (8)$ & $0.09023(48)$ & $316$ & $188$ & $396$ \\
      $48^3 \times 144$ & $0.0582 (4)$ & $0.05926(33)$ & $329$ & $186$ & $386$  \\
      $64^3 \times 192$ & $0.0441 (2)$ & $0.04406(27)$ & $315$ & $142$ & $400$  \\
      $96^3 \times 288$ & $-$ & $0.03271(20)$ & $309$ & $-$ & $451$ \\ 
      $40^3 \times 64$  & $0.1189 (10)$ & $0.12225(64)$ & $216$ & $100$ & $200$  \\
     \hline\hline
    \end{tabular}
    \caption{Details of the lattice QCD ensembles utilized for calculations with Overlap and HISQ valence quarks. The table lists the lattice volume $N_s^3 \times N_t$, the corresponding lattice spacings for Overlap~\cite{MILC:2012znn,MILC:2015tqx,PhysRevD.98.074512} and HISQ~\cite{HPQCD_2020} actions, the sea-pion mass $M_{\pi}^{\text{sea}}$ (MeV)~\cite{PhysRevD.98.074512}, and the number of measurements $n_{\text{meas}}$ used for Overlap and HISQ calculations.}
    \label{tab:Lattice_details_ov_hisq}
\end{table}

\subsection{Valence Operators: Overlap and HISQ}
The Overlap valence action offers several advantages for this study: it is free from $\mathcal{O}(a)$ discretization errors, enabling cleaner continuum extrapolations; it provides improved control over operator mixing; and its symmetry properties make it particularly effective for heavy-quark systems such as the $\Omega_{ccc}$ (also demonstrated in several of our earlier works). To build the interpolating operators for the $\Omega_{ccc}$, we follow the symmetry-guided construction of Ref.~\cite{Basak:2005ir}. Because the baryon is color antisymmetric and trivially symmetric in flavor, the spin component must be fully symmetric, leading to a uniquely allowed local state with quantum numbers $J^{P}=3/2^{+}$. On the lattice, this state belongs to the $H^{+}$ irreducible representation of the cubic group \cite{Johnson:1982yq}. Using four-component Overlap spinors, this irrep can be realized through two independent operator embeddings, which we include in our variational basis. Further discussion of this construction can be found in our detailed $\Omega_{ccc}$ study \cite{Dhindsa:2024erk}, as well as in our study of fully charmed and strange dibaryon systems \cite{Dhindsa:2025gae}, where the same strategy has been employed. We analyse the resulting correlation matrices variationally using the generalized eigenvalue problem (GEVP). In addition to the usual forward-propagating correlators, we also include the time-reversed backward-propagating correlators in our study.

The HISQ valence action enhances internal consistency and eliminates mixed-action effects that could otherwise complicate the continuum limit. Although operator mixing can be nontrivial for staggered formulations, such issues are minimal for single-flavor systems like the $\Omega_{ccc}$, making HISQ a reliable complementary valence discretization in our study. Further details on the smearing procedures and the specific terms entering the HISQ action can be found in Ref.~\cite{Dhindsa:2024erk}. The corresponding interpolating operator takes the form \cite{GoltermanBaryon}:
\begin{eqnarray}
    \mathcal{O}_{\Omega_{ccc}}(t) =  \epsilon_{abc} D_1c^a(\mathbf{x},t)D_2c^b(\mathbf{x},t)D_3c^c(\mathbf{x},t) \; .\label{OmegaOp}
\end{eqnarray}
Here, $c^{a}(\mathbf{x},t)$ denotes the staggered charm quark field, with 
$a$, $b$, and $c$ labeling color indices. The spatial derivative operators 
$D_i$ act through a symmetric nearest–neighbour shift, $D_i\, c^{a}(\mathbf{x},t)
    = \tfrac12 \big[ c^{a}(\mathbf{x}+\hat{i},t)
    + c^{a}(\mathbf{x}-\hat{i},t) \big].$
The resulting interpolating operator transforms in the $8'$ irreducible 
representation of the geometric time-slice (GTS) group, and corresponds to 
the $A_{2}^{-}$ irrep of the octahedral group $\mathcal{O}_h$ 
\cite{GoltermanBaryon, Bailey:2006zn}. We study the exponential decay of the resulting two-point correlation functions constructed from these operators, employing both single- and double-exponential fit forms to robustly extract the ground-state energy.

In both the Overlap and HISQ calculations, we employ a wall-source point-sink setup. 
For HISQ correlators, the staggered formulation induces time oscillating contributions from opposite-parity states; to suppress these oscillations, we smooth the correlators by forming multiplicative combinations of correlator values at several temporal shifts. These smoothed correlators, which remove the alternating phase, are then used for the subsequent spectral analysis. Using the Overlap data, the correlation matrices were also analyzed through a transverse GEVP, confirming that contributions from excited states are negligible \cite{Chakraborty:2024exj}.

\subsection{Mass Spectra and Splittings}
\begin{figure}[h!]
    \centering
    \includegraphics[scale=0.24]{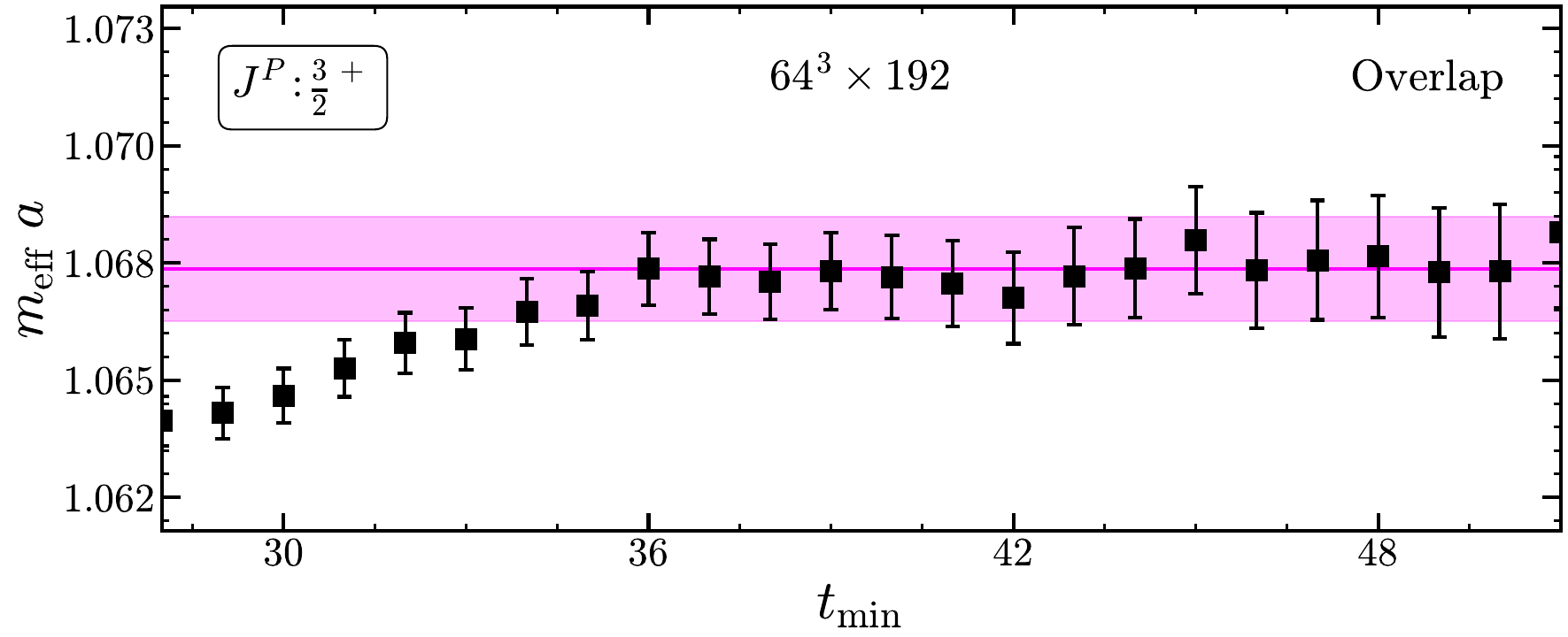} 
    \includegraphics[scale=0.24]{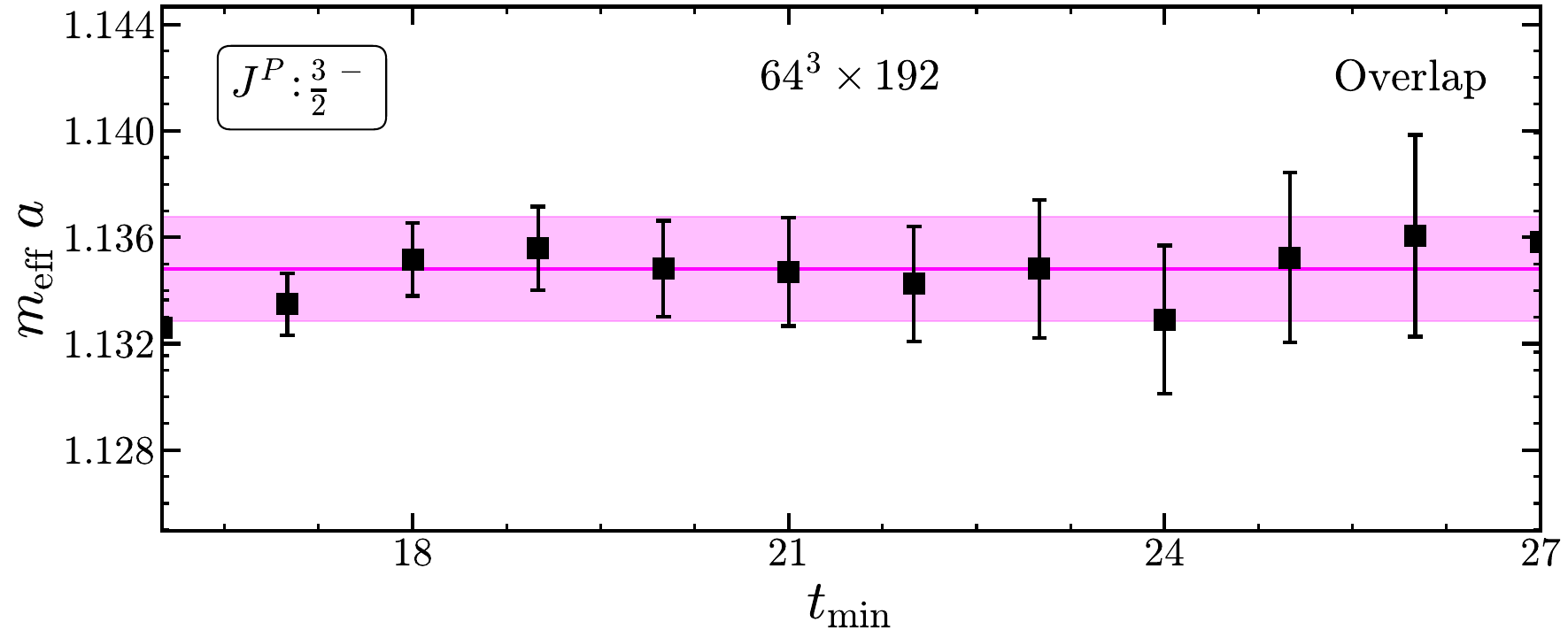}
    \includegraphics[scale=0.24]{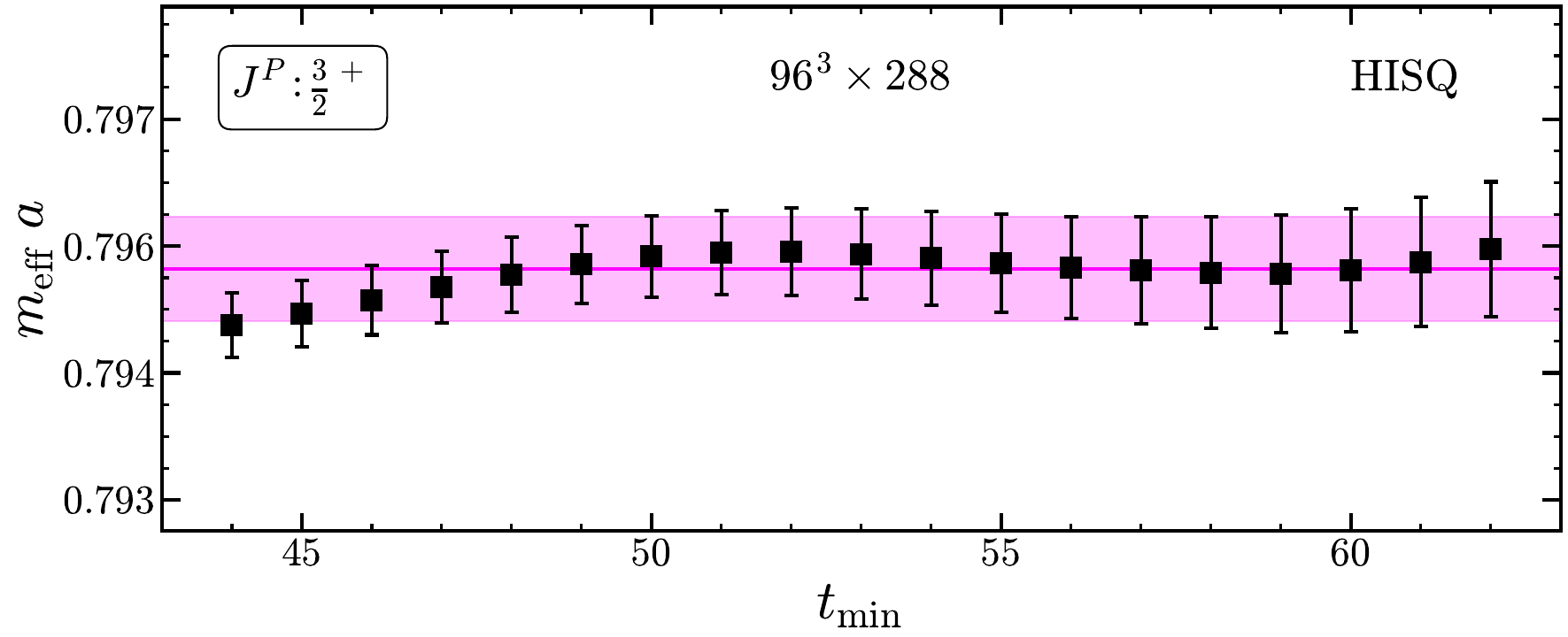} 
    \includegraphics[scale=0.24]{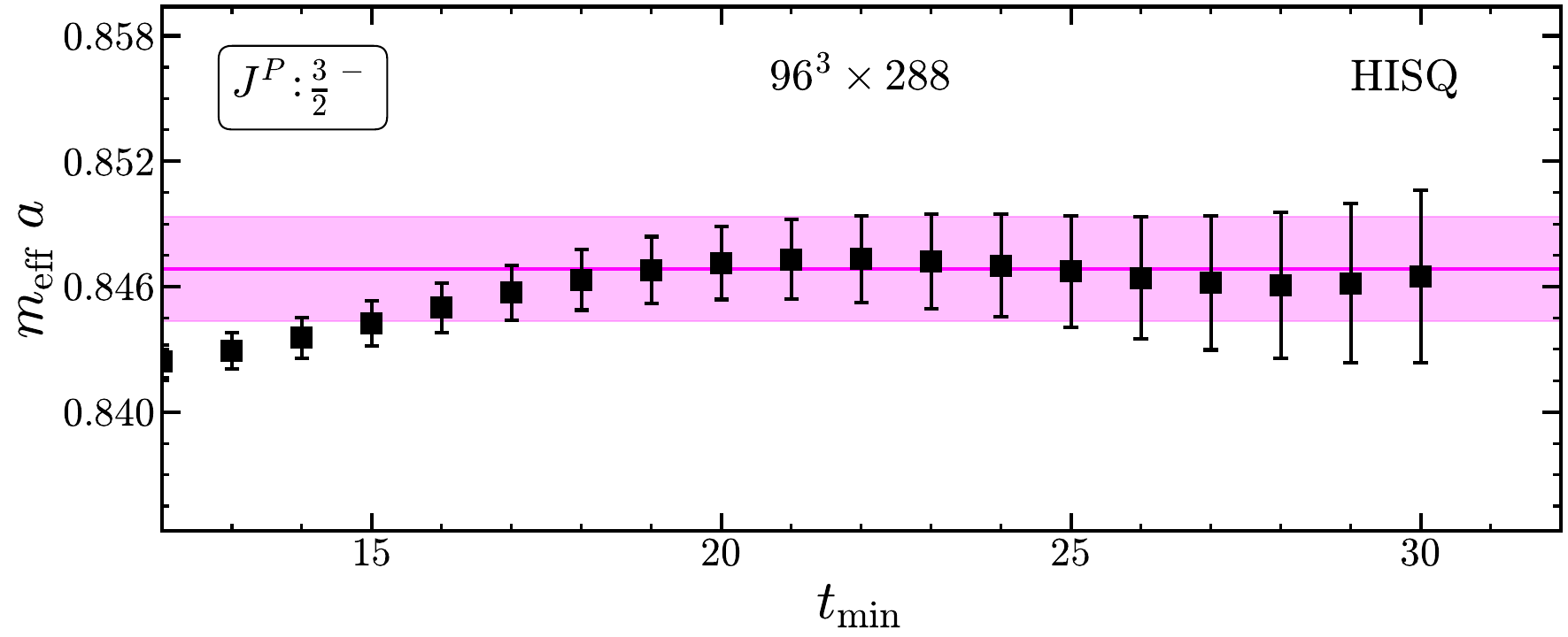}
    \caption{Dependence of the extracted mass values (in lattice units) on the choice of $t_{\min}$, on the $64^3 \times 192$ Overlap ensemble and the $96^3 \times 288$ HISQ ensemble.}
    \label{fig:tminmax}
\end{figure}

The signal quality is illustrated in Fig.~\ref{fig:tminmax} through the $t_{\min}$ dependence of the extracted masses for the finest ensembles used in the Overlap and HISQ analyses for both the positive- and negative-parity states. The shaded band in Fig.~\ref{fig:tminmax} represents the statistical and systematic uncertainty associated with the choice of fit range. A similar fit-window stability analysis was performed across all remaining ensembles to ensure uniform control over systematics.

\begin{figure}[h!]
    \centering
    \includegraphics[scale=0.29]{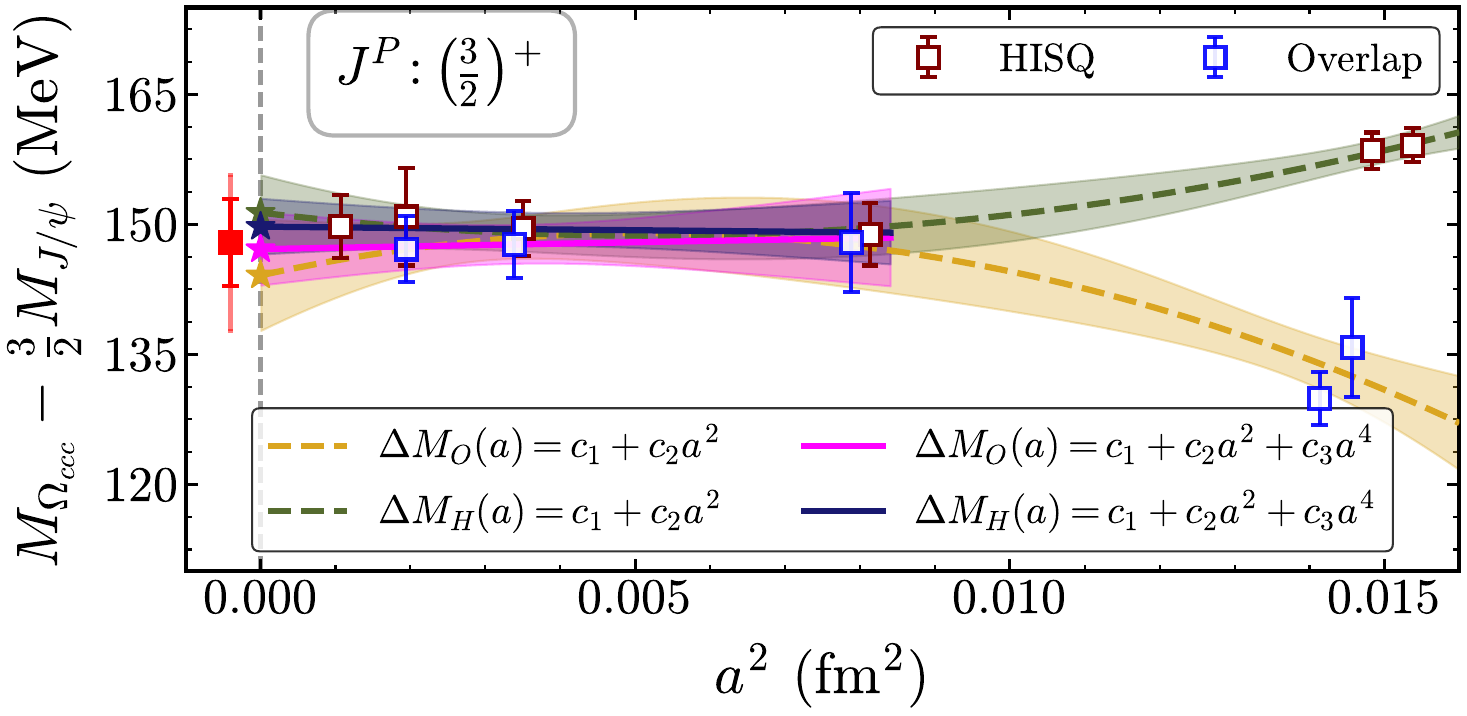} ~~
    \includegraphics[scale=0.29]{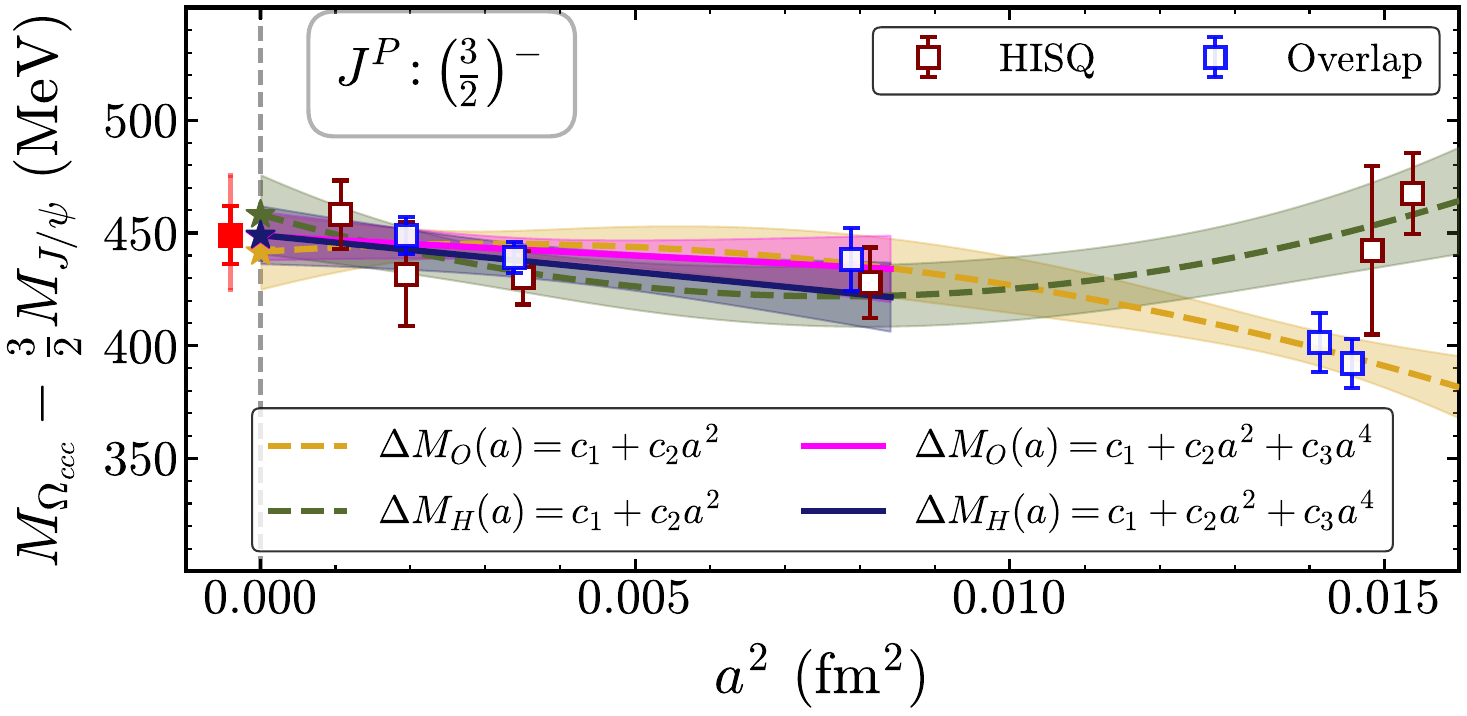}
    \caption{Continuum extrapolation of 
$\Delta M_{\Omega_{ccc}}$  for the 
positive (left) and negative (right) parity channels. Different fit forms are displayed with the color codes indicated in the legends. The shaded bands denote the $1\sigma$ uncertainties obtained from bootstrap resampling. The stars mark the corresponding continuum limits, while the red filled squares indicate the final estimates, obtained from a symmetrized average of the linear extrapolations from the Overlap and HISQ data; the enlarged error bars incorporate the spread across the alternative fit forms.}
    \label{fig:split}
\end{figure}

Being composed entirely of heavy quarks, triply charmed baryons are particularly sensitive to lattice discretization effects. The large bare charm mass enhances $\mathcal{O}(ma^2)$ artifacts, making the absolute mass values less reliable as standalone observables. To mitigate these heavy-quark discretization effects, we base our conclusions primarily on mass splittings rather than the raw effective masses, as splittings are significantly less sensitive to such systematic uncertainties. To reduce heavy-quark discretization effects, we work with the subtracted mass
\begin{equation}
    a\Delta M_{\Omega_{ccc}} =
    aM^{L}_{\Omega_{ccc}} - \tfrac{3}{2}\,aM^{L}_{J/\psi},
\end{equation}
where the reference mass removes the dominant valence charm contribution.  
Although one can alternatively use the spin-averaged ${\overline{1S}}$ charmonium energy, the $J/\psi$ mass is generally reproduced more accurately on our ensembles, making it the preferred choice for the splittings shown here.  
A full discussion of both subtraction schemes and the corresponding continuum extrapolations can be found in Ref.~\cite{Dhindsa:2024erk}. In Fig.~\ref{fig:split}, we present the continuum extrapolation of the mass splittings 
$\Delta M_{\Omega_{ccc}}$ in physical units for both parity channels. 
Two forms of cutoff dependence are considered: 
(i) $\Delta M(a) = c_1 + c_2 a^2$, and 
(ii) $\Delta M(a) = c_1 + c_2 a^2 + c_3 a^4$. 
The marker and color conventions are explained in the figure and its caption.

Once the continuum-extrapolated value of $\Delta M^{\rm cont}_{\Omega_{ccc}}$ is obtained, the physical mass of the baryon is determined by adding the corresponding valence charm contribution from experiment as $
M^{\rm phys}_{\Omega_{ccc}} = \Delta M^{\rm cont}_{\Omega_{ccc}} + \frac{3}{2} M^{\rm exp}_{J/\psi}$. A clear agreement is observed between the mass estimates obtained using the Overlap and HISQ valence actions on the finer lattices, resulting in a consistent continuum-extrapolated value. Our final estimates are: $ M_{\Omega_{ccc}(3/2^+)} = 4793~\text{MeV}, M_{\Omega_{ccc}(3/2^-)} = 5094~\text{MeV}$.

In addition to the analysis discussed, we verified that the hyperfine splitting in the continuum limit agrees with the experimental value for both valence actions used.  
For the HISQ ensembles, alternative fit forms, such as 
$c_1 + c_2 a^2 + c_5 \alpha_s(1/a) (m_c a)^2$ and 
$c_1 + c_4 a^4 + c_5 \alpha_s(1/a) (m_c a)^2$, 
were also tested and yielded consistent results.  
Moreover, the continuum splitting between the $3/2^+$ and $3/2^-$ states, as well as the continuum fits of these splittings on each lattice, are in good agreement. 
A detailed discussion of these checks and results is provided in Ref.~\cite{Dhindsa:2024erk}. 

\subsection{Error Budget}
For a precise prediction that can reliably guide experimental searches, it is crucial to quantify all sources of uncertainty. In our calculation, we consider both statistical and systematic errors. The systematic uncertainties arise from several sources, including discretization effects estimated by comparing different fit forms, the Overlap and HISQ actions scale setting, tuning of the charm quark mass, unphysical sea quark masses requiring chiral extrapolation, taste-splitting effects specific to HISQ, mixed-action effects for Overlap valence on HISQ sea, finite-volume effects, and perturbative estimates of electromagnetic corrections for both $\Omega_{ccc}$ and the $J/\psi$. Further discussion of the various sources of uncertainty can be found in Ref.~\cite{Dhindsa:2024erk}.
 After combining all these contributions, we obtain the final estimates for the ground-state masses as: 
$M_{\Omega_{ccc}(3/2^+)} = 4793(5)\left(^{+11}_{-8}\right) ~\text{MeV}$ and 
$M_{\Omega_{ccc}(3/2^-)} = 5094(12)\left(^{+19}_{-17}\right) ~\text{MeV}$.

\section{Conclusion}
We investigated the masses of the $\Omega_{ccc}$ baryon for both parities using lattice QCD with five lattice spacings, two different volumes, two valence quark actions, and several continuum-extrapolation strategies. This work provides the most precise and systematically controlled determination of the $\Omega_{ccc}$ ground-state masses to date, establishing a reliable benchmark for future experimental studies.


\acknowledgments
This work is supported by the Department of Atomic Energy, Government of India, under Project Identification Number RTI 4002. Computations were carried out on the Cray-XC30 of ILGTI, TIFR (which has recently been closed), and the computing clusters at the Department of Theoretical Physics, TIFR, Mumbai, and IMSc Chennai. We are thankful to the MILC collaboration and, in particular, to S. Gottlieb for providing us with the HISQ lattice ensembles. We would also like to thank  Ajay Salve, Kapil Ghadiali, and T. Chandramohan for computational support. NM and DC also thank Amol Dighe for discussions.  MP gratefully acknowledges
support from the Department of Science and Technology, India, SERB Start-up Research Grant No. SRG/2023/001235.


\bibliographystyle{JHEP}
\bibliography{proc}
\end{document}